\documentclass[a4paper]{jpconf}
\usepackage{graphicx}
\begin{document}
\title{A procedure to characterize the performance of Energy-Resolved Detectors (EDX)}

\author{F.J.~Iguaz, T.~Saleem\footnote[1]{Present address: WEEROC, 4 avenue de la Baltique, BP 515, 91140 Villebon sur Yvette, France} and N.~Goyal}

\address{Synchrotron SOLEIL - Detectors Group, L'Orme des Merisiers, Departamentale 128, 91190 Saint Aubin, France}

\ead{francisco-jose.iguaz-gutierrez@synchrotron-soleil.fr}

\begin{abstract}
The detector group of Synchrotron SOLEIL is monitoring the performance of Energy-Resolved Detectors (EDX) and their associated electronics since last five years. A characterization procedure has been developed for this purpose and for Site Acceptance Tests (SATs) of new EDXs installed at beamlines. This manuscript presents the procedure, illustrating it with an example.
\end{abstract}

\section{Introduction}
\label{sec:intro}
Energy-Resolved Detectors (EDX), either Silicon Drift Detectors (SDD) or High Purity Germanium (HPGe), equipped with a Digital Pulse Processor (DPP), are detection systems commonly used in synchrotron experiments like X-ray Fluorescence (XRF) and X-ray Absorption Spectroscopy (XAS). These detectors are particularly useful if the element concentration is very low or if the sample is very thick.

SOLEIL beamlines have 35~EDXs and 24~DPPs in operation. EDXs are mainly SDDs (32) because beamline energy ranges are within SDD maximum efficiency (1-20~keV). Meanwhile, 2~HPGe and 1~CdTe are operated at beamlines with the highest energy range (PSICHE and SAMBA). EDXs normally fail due to material aging or accidental misuse. As failures may not be evident during beamline operation, beamline scientists normally contact detectors group to monitor the EDX's heath. A characterization procedure has been created for this purpose and for Site Acceptance Tests (SAT) of new EDXs installed at beamlines.

The procedure is composed of four steps:
\begin{enumerate}
 \item The measurement of the dynamic range and reset period by an oscilloscope, which provides an estimation of the leakage current. This parameter is a good indicator of the sensor status.
 \item The acquisition of 100k signal waveforms to calculate the preamplifier gain and the risetime distribution. The gain is defined by the constructor and any variation could indicate an electronics problem. The risetime distribution is related to the charge diffusion in the sensor and any degradation could indicate a degradation in vaccum conditions.
 \item The acquisition of energy spectrum of a $^{55}$Fe source at low ICR ($\sim$10~kcps) with a XIA-XMAP~\cite{Hubbard1996} or DANTE~\cite{Iguaz:2023otz} DPP at different peaking time values. The optimum peaking time is defined by the type of preamplifier (CUBE or JFET) and the sensor capacitance ($\sim$fF for SDDs, $\sim$1~pF for HPGe), while the optimum energy resolution at 5.9~keV is defined by the sensor (size, entrance window) and the type of preamplifier. A change in this dependence indicate a problem (grounding, preamplification, sensor, etc).
 \item The acquisition of energy spectrum of a $^{55}$Fe source with a Xspress3X/M DPP~\cite{FARROW1995567} at ICR between 10 and 1000~kcps to measure the energy resolution, the peak stability, the dead time and the pile-up rejection factor. These performance parameters provide information to create a calibration settings file adapted to the beamline.
\end{enumerate}

The full procedure is used for new detectors, while the three first steps are followed for monitoring the detector peformance. This procedure is illustrated here by a MIRION X-PIPS detector (13 $\times$ 80~mm$^2$ collimated area), in operation at MARS and SAMBA beamlines since June 2020. A photo of the SDD 13EM is shown in Fig.~\ref{fig:edx_sdd_range} (left).

\section{Dynamic range and reset period}
\label{sec:edx_sdd_osc}
\begin{figure}[htb!]
\centering
\includegraphics[width=.46\textwidth]{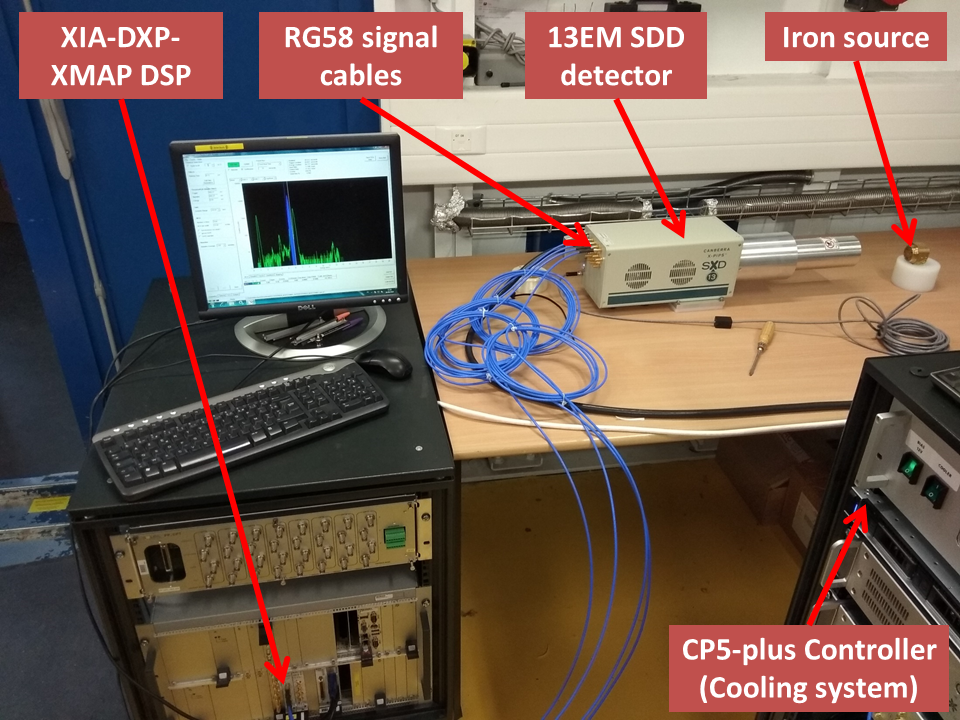}
\includegraphics[width=.51\textwidth]{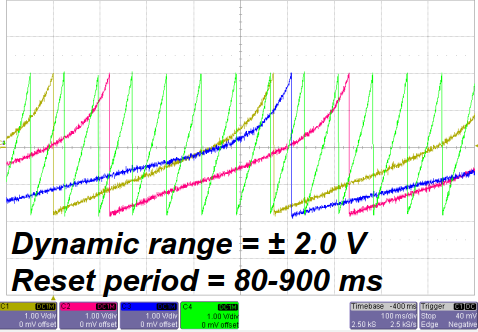}
\caption{Left: Photo of the SDD 1EM during its performance test made in July 2020. Right: Preamplifier output signal of channels 1 to 4 in absence of X-rays. Note that the channel 4 shows a shorter reset period (76~ms) compared to others (460, 510 and 710~ms).}
\label{fig:edx_sdd_range}
\end{figure}

The dynamic range and the reset period in absence of x-rays radiation were first measured for all pixels, connecting them one by one to the oscilloscope. Two or three wavefors were recorded in each case. One example is shown in Fig.~\ref{fig:edx_sdd_range}. The dyanmic range is $\pm$2~V and the RESET period in absence of X-rays radiation is 76-890~ms. The elements work in asynchronous reset mode. The leakage current values of sensors are between 0.05 and 0.6~pA. These values are lower than $\sim$2~pA, which is well bellow the value of $\sim$2~pA, where the energy resolution starts to degrade.

\section{Preamplifier gain and signal risetime}
\begin{figure}[htb!]
\centering
\includegraphics[width=.49\textwidth]{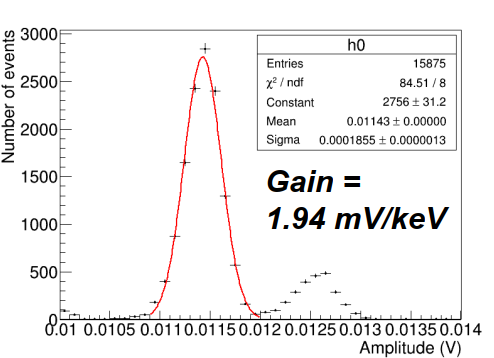}
\includegraphics[width=.49\textwidth]{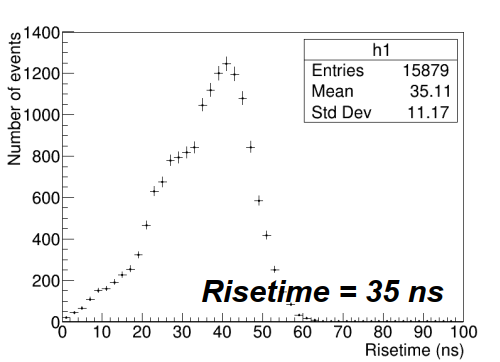}
\caption{Amplitude (left) and risetime distributions (right) of induced signals by $^{55}$Fe source.}
\label{fig:edx_sdd_gainrt}
\end{figure}

To acquire $\sim$100k waveforms for each channel using an oscilloscope, we place a $^{55}$Fe source in front of the detector. Waveforms were analysed offline to calculate the amplitude and risetime (10-90) distributions for all pixels, as illustrated in Fig.~\ref{fig:edx_sdd_gainrt} for pixel 1. For each amplitude distribution, the x-ray line corresponding to 5.9~keV x-rays was fitted by a Gaussian function to estimate precisely its mean amplitude and the preamplifier gain, calculated as the ratio of the mean amplitude in mV and 5.9~keV. Gain settings are around 2~mV/keV, as defined by the constructor. The risetime distribution has a mean value lower than 50~ns, so as the element could show a good performance a high ICR, i.e., an equivalent shorter PT.

During the detector operation in 2021, an increase of the cooler power was observed (0.4~Watts/day), which was caused by a leak at the Beryllium window. This problem in vacuum side was also visible at signal risetime distribution as the mean values increased from 35 to 50~ms. The problem was solved by fixing the leak with epoxy and the outgassing of the molecular sieves.

\section{Performance at low X-ray flux}
\label{sec:edx_sdd_xmap}
\begin{figure}[htb!]
\centering
\includegraphics[width=.49\textwidth]{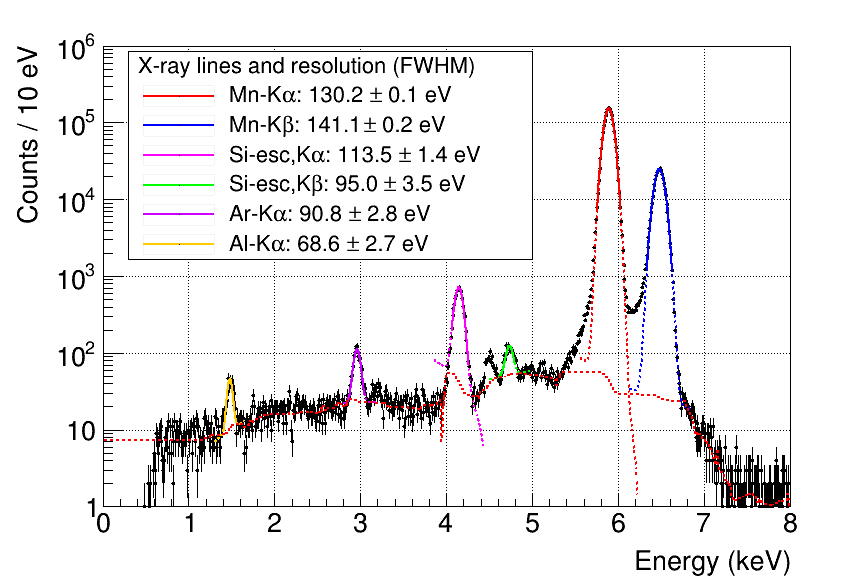}
\includegraphics[width=.49\textwidth]{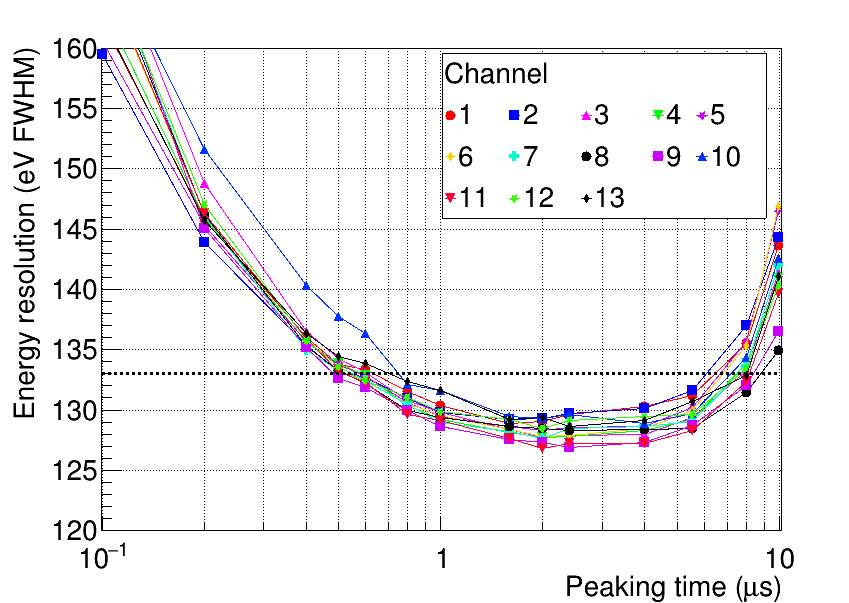}
\caption{Left: Energy spectrum of the iron source for the channel 1 of the SDD 13EM. The different x-ray lines have been identified and fit by a Gaussian function, after having
removed the Compton background (dashed red line). The energy resolution (FHWM) is given for each case.
Right: Dependence of the energy resolution at 5.9 keV with the peaking time for all channels.}
\label{fig:edx_sdd_specpt}
\end{figure}

The detector performance was characterized in terms of energy resolution, peak-to-background (PTB) and peak-to-tail (PTT), at low ICR with the help of a XIA-XMAP board and an iron source. Channels were successively connected to the four channels of the DPP in batches of four (as shown in Fig.~\ref{fig:edx_sdd_range}, left) by four RG58 signal cables. The XIA-XMAP board was calibrated using the x-rays of the iron source. Energy spectra of 30 seconds were acquired at an ICR of 10~kcps, a threshold of 500~eV and for peaking time (PT) values varying from 0.1 to 10~\textmu s. A long spectrum of 240 seconds was also acquired for all channels at the optimum PT (2~\textmu s) to measure precisely the background levels at 2 chosen values of 5.3~keV and 1.0~keV, later used to calculate PTB and PTT values. An example of one spectrum is shown in Fig.~\ref{fig:edx_sdd_specpt} (left).

Each energy spectrum is composed of two intense manganese fluorescence X-ray lines (K$_\alpha$-line at 5.9~keV and K$_\beta$-line at 6.4~keV), a continuous Compton/Rayleigh scattering background, two escape peaks situated at a lower energy from main lines (1.74~keV less, at 4.16 and 4.66~keV) and generated by the silicon sensor, argon fluorescence X-ray line from air (K$_\alpha$-line at 2.96~keV), aluminium fluorescence X-ray line from collimator (at 1.5~keV). In the offline analysis of each spectrum based on ROOT~\cite{Brun:1997pa}, the continuous background has been fitted and substracted, and X-ray lines have been identified and adjusted by a Gaussian function to estimate the peak position (mean value) and energy resolution (FWHM value). For the manganese K$_\alpha$-line, the statistical error of energy resolution is less than 0.1\% and is negligible for peak position.

The dependence of the energy resolution at 5.9 keV with the peaking time is shown in Fig.~\ref{fig:edx_sdd_specpt} (right). The shape is defined by the three noise components: current noise (long PT, right side), voltage noise (short PT, left side) and Flicker noise (minimum value, central part)~\cite{Spieler:2005si}. The optimum energy resolution (127-129~eV) is within constructor values and near the Fano limit (121~eV~\cite{PEROTTI1999356}), while the optimum PT is near 1~\textmu~s, the expected value for CUBE preamplifier~\cite{6154396}. The measured PTT and PTB values are 1.4-2.0 $\times 10^4$, as expected for a SDD.

\section{Performance under high X-ray flux}
\label{sec:edx_sdd_xspress3}
The detector performance in terms of energy resolution, peak stability, dead time and pile-up rejection was characterized at high ICR with a 13 channel Xspress3X DPP in a beam test at SAMBA beamline. A manganese foil, rotated 45~degress with respect to the detector and the beam axis, was installed. The beam energy was fixed to 10~keV, the beam flux was $\sim 10^{10}$~ph/sec and the detector-to-foil sample was varied to get ICR values between 10 and 6000~kcps.

\begin{figure}[htb!]
\centering
\includegraphics[width=.60\textwidth]{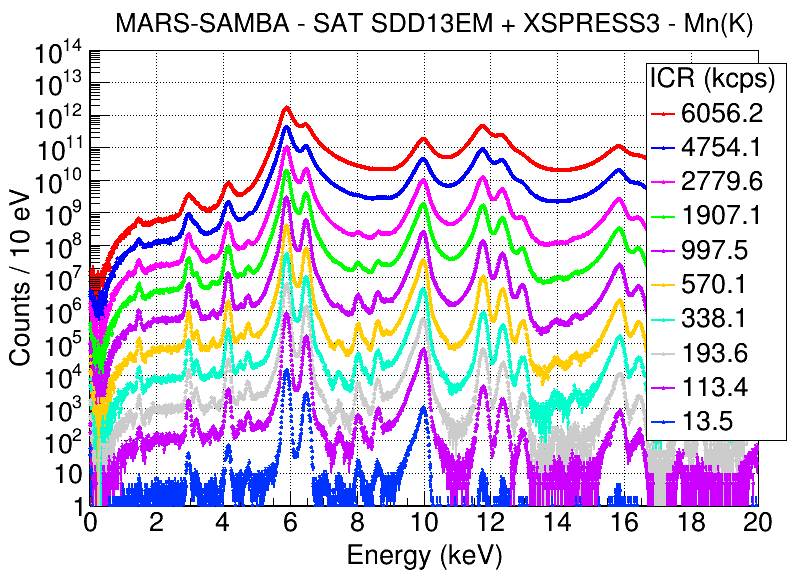}
\caption{Energy spectrum of the iron source at different ICR values for the channel 1 of SDD 13EM measured by Xspress3 DPP.}
\label{fig:edx_sdd_x3_spec}
\end{figure}

The energy spectra of channel 1 are shown in Fig.~\ref{fig:edx_sdd_x3_spec}. Each spectrum shows the elastic peak at 10.0~keV, manganese fluorescence lines (at 5.9 and 6.49~keV), their corresponding escape peaks (at 4.15 and 4.75~keV), argon lines (at 2.96 and 3.19~keV), aluminium lines (at 1.5~keV) and several pile-up lines at energies higher than 10~keV. At high photon flux, the energy resolution degrades and the intensity of pile-up lines increases due to DPP pile-up rejection value.

\begin{figure}[htb!]
\centering
\includegraphics[width=.49\textwidth]{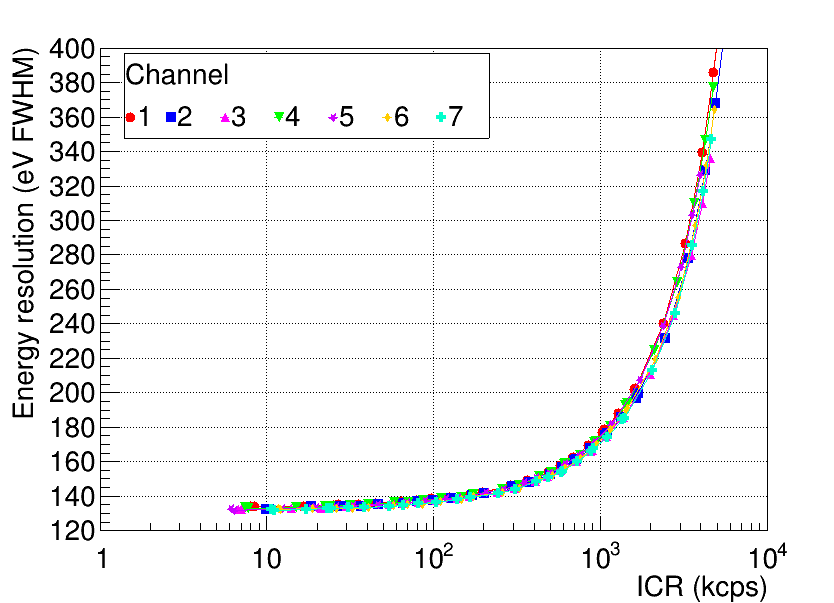}
\includegraphics[width=.49\textwidth]{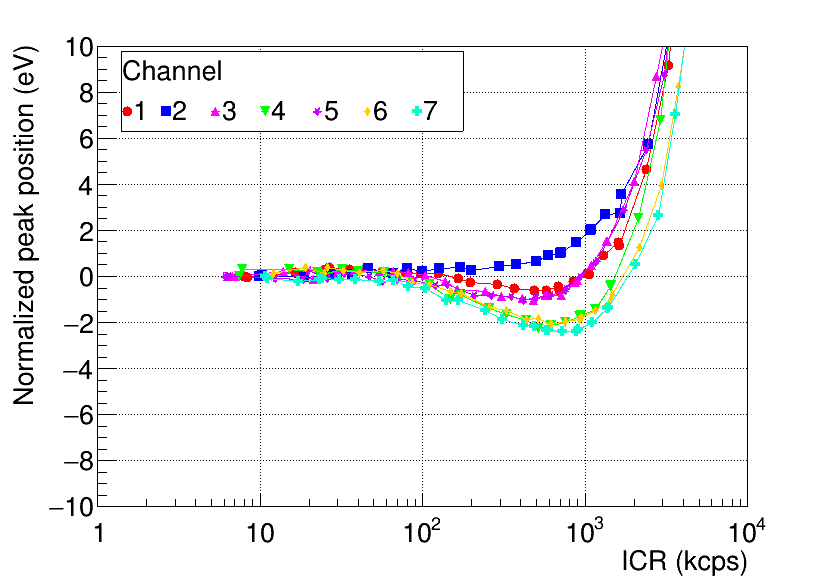}
\caption{Dependence of the energy resolution (FWHM) at 5.9~keV (left) and the normalized peak position (right) with the ICR for seven channels measured by Xspress3 DPP.}
\label{fig:edx_sdd_x3_resicr}
\end{figure}

The dependence of the energy resolution at 5.9~keV with the ICR is shown in Fig.~\ref{fig:edx_sdd_x3_resicr} (left). The measured values at 500~kcps (150-157~eV), 1~Mcps (170-177~eV) and 2~Mcps (210-220~eV) fulfil the criteria for energy resolution at high ICR, as their are less than 175~eV at 500~kcps, 200~eV at 1~Mcps and 230~eV at 2~Mcps. The dependence of the peak position with the ICR is shown in Fig.~\ref{fig:edx_sdd_x3_resicr} (right). The peak shift at 1~Mcps ($\pm$2.0~eV) is almost negligible.

\begin{figure}[htb!]
\centering
\includegraphics[width=.49\textwidth]{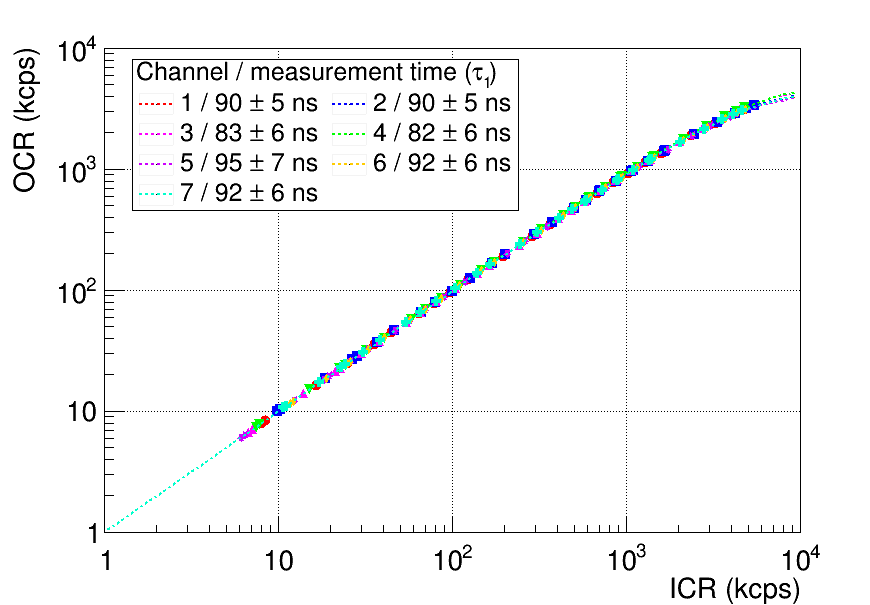}
\includegraphics[width=.49\textwidth]{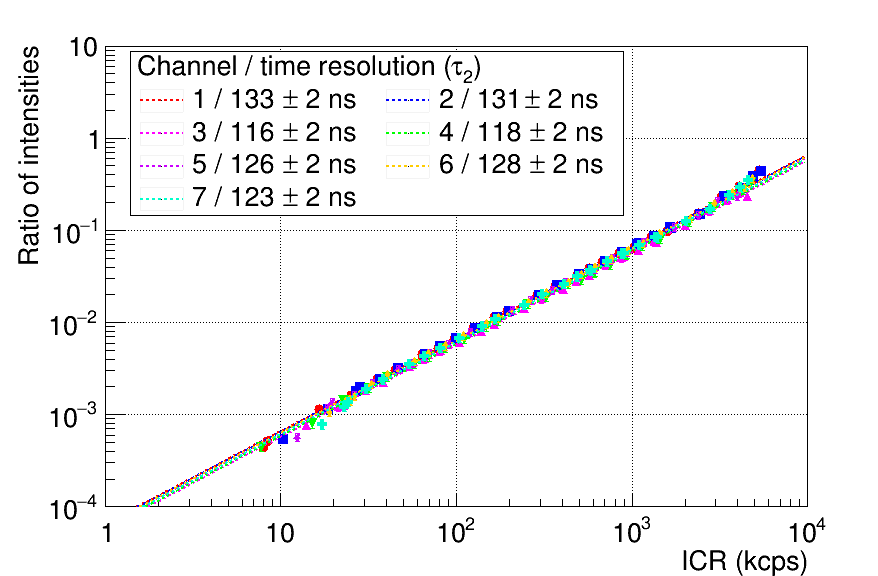}
\caption{Dependence of the OCR (left) and the peak ratio (right) with the ICR for seven channels measured by Xspress3 DPP.}
\label{fig:edx_sdd_x3_ocrint}
\end{figure}

The dead time and pile-up rejection factors were charaterized by the measurement time and the time resolution, following the reference~\cite{Bordessoule2019}.
\begin{itemize}
    \item The measurement time ($\tau_1$) is defined as the minimum interval time between two successive impulses that  is acceptable by the DPP for a correct reconstruction of energy values. This parameter is calculated from the dependence of OCR and ICR, shown in Fig.~\ref{fig:edx_sdd_x3_ocrint} (left), using the expression:
\begin{equation}
 OCR = ICR \times \exp(-\tau_1 \times ICR)
\label{eq:edx_sdd_tau1}
\end{equation}
    The estimated values for $\tau_1$ (83-92~ns) are not far from other tests of the Xspress3M with a SDD equipped with CUBE preamplifier ($\sim$75~ns). The equivalent dead time values at 1 and 2~Mcps (5\% and 35\%) are within specifications (10\% and 50\%).
    \item The time resolution ($\tau_2$, or pile-up rejection power), is defined as the minimum interval time between two successive impulses for which the DPP can separate them. This parameter is estimated from the dependence of the pile-up ratio of intensities (i.e., the ratio of intensities of the K$_\alpha$-K$_\alpha$ pile-up line ($Int(2)$) and the $K_\alpha$ line ($Int(1)$) with the ICR, shown in Fig.~\ref{fig:edx_sdd_x3_ocrint} (right), using the expression
\begin{equation}
 \frac{Int(2)}{Int(1)} = ICR \times \tau_2 / 2
 \label{eq:edx_sdd_tau2}
\end{equation}
    The estimated values for $\tau_2$ (116-136~ns) are not far from other tests of the Xspress3M with a SDD equipped with CUBE preamplifier ($\sim$100~ns). The pile-up intensity at 1~Mcps (0.58-0.68\%) are within specifications (1\%).
\end{itemize}

\section{Conclusions}
\label{sec:conc}
A procedure to characterize the performance of EDX detector has been presented. The procedure quantifies the sensor leakage current, the pre-amplifier gain and risetime distribution, the detector noise at low X-ray flux and the ultimate performance at high X-ray flux, in terms of energy resolution, peak stability, dead time and pile-up rejection power. Based on these results, customized calibration files are generated and adapted to the specific beamlines requirements, improving their experimental results.

\ack
The authors would like to thank beamline scientists of Synchrotron SOLEIL, specially Emiliano Fonda and Nicolas Guignot from SAMBA and PSICHE beamlines, for their support in creating this procedure.

\section*{References}
\bibliographystyle{JHEP}
\bibliography{20240830_FJIguaz_SRI2024}
\end{document}